\documentclass[aps,pra,twocolumn,floatfix,groupedaddress,footinbib,notitlepage,showpacs]{revtex4-1}
\usepackage{graphicx,graphics,times,bm,bbm,amssymb,amsmath,amsfonts,color}
\newcommand{\ket}[1]{|#1\rangle}
\newcommand{\bra}[1]{\langle#1|}

\newcommand{\ketbra}[2]{|#1\rangle\langle #2|}
\begin{document}
\title{Quantum discord and non-Markovianity of quantum dynamics}
\author{S. Alipour}
\affiliation{Department of Physics, Sharif University of Technology, Tehran, Iran}
\author{A. Mani}
\affiliation{Department of Physics, Sharif University of Technology, Tehran, Iran}
\author{A. T. Rezakhani}
\affiliation{Department of Physics, Sharif University of Technology, Tehran, Iran}

\begin{abstract}
The problem of recognizing (non-)Markovianity of a quantum dynamics is revisited through analyzing quantum correlations. We argue that instantaneously-vanishing quantum discord provides a necessary and sufficient condition for Markovianity of a quantum map. This is used to introduce a measure of non-Markovianity. This measure, however, requires demanding knowledge about the system and the environment. By using a quantum correlation monogamy property and an ancillary system, we propose a simplified measure with less requirements. Non-Markovianity is thereby decided by quantum state tomography of the system and the ancilla.
\end{abstract}

\maketitle

\section{Introduction}

Real quantum systems are inevitably open because of interaction with their ambient environment. As a result, information, in principle, can leak into the environment, and may be fed back later in a different form into the system \cite{Breuerbook}. This scenario makes accurate description of the related dynamics difficult \cite{Cubitt-2}, because one often has no much information/control about/over the environment---unless, e.g., the environment can be controllably engineered \cite{Verstraete-engineering}. Nevertheless, under some specific (ideal) conditions, such as weak coupling with a memoryless environment and Born-Markov approximation, the underlying system evolution can be well described by a quantum \textit{Markovian} dynamical equation \cite{Breuerbook} (see also Ref.~\cite{Lidar-chem} for an alternative derivation). This approximation has been proved useful in various situations, such as devising quantum error-correction schemes \cite{Nielsen:book}.

In this case, the dynamics of the system for any \textit{interval} $\tau$---regardless of the start time $t_0$---is given by a completely-positive (CP) map $\widehat{\mathcal{E}}(\tau)$, having the ``dynamical semigroup" property $\widehat{\mathcal{E}}(\tau)  \widehat{\mathcal{E}} (\tau') = \widehat{\mathcal{E}}(\tau+\tau')$, for any $\tau,\tau'\geq0$ \cite{cirac:Assessing NM, Cirac:Dividing quantum channels}; or equivalently, the dynamical equation $d\varrho(\tau)/d\tau = \widehat{\mathcal{L}}\bigl[\varrho(\tau)\bigr]$ governs the system density matrix, where the time-independent $\widehat{\mathcal{L}}$ has the following Lindblad form:
\begin{eqnarray}
\widehat{\mathcal{L}}[\cdot]= -i[H,\cdot]+\sum_{m} r_{m}\bigl(2 F_m \cdot F_m^{\dag}-\{F_m^{\dag}F_m,\cdot\}_+\bigr),
\label{LF}
\end{eqnarray}
leading to $\widehat{\mathcal{E}}(\tau)\equiv e^{\tau\widehat{\mathcal{L}}}$. Here, $r_m\geq 0~\forall m$, $H$ is a Hermitian operator, and $F_m$s are some operators acting on the system's Hilbert space \cite{Breuerbook,Lindblad}. Note that the case of time-dependent $r_m$s is also referred to as a time-dependent Markovian evolution when $r_m(t)\geq0~\forall t$ \cite{Breuerbook}.

This simplified picture (even its beyond-Markovian version---Redfield theory), although fairly well applicable to various situations, often fails to capture dynamics of complex quantum manybody systems in condensed matter or in some biological complexes (which may feature quantum effects). In such cases, at least one of the assumptions of the Born-Markov (or Redfield) theory breaks down \cite{Ishizaki-Fleming:09,Cheng-Fleming:09, c2}.

These difficulties have spurred introducing various formalisms to somehow incorporate non-Markovian effects into open system's dynamics, e.g., through adding memory kernels \cite{SL-pM} or even devising specific dynamical equations valid for non-Markovian regimes (see, e.g., Refs.~\cite{Breuerbook,NM-papers}).

Although deciding whether a given quantum channel/dynamics is Markovian (or not) is in complex-theoretic sense hard \cite{Cubitt}, there exist a number of measures of Markovianity based on the Lindblad form \cite{cirac:Assessing NM,Andersson}. In addition, some experimentally measurable criteria have been introduced for detecting non-Markovianity of an evolution by using, e.g., increasing (decreasing) the distance (fidelity) of two density matrices in time \cite{Breuer,indian}, or increasing the entanglement between the system and an isolated ancilla in some instances of time \cite{Rivas}. These measures provide sufficient conditions for a channel to be non-Markovian, which leaves open search for more delicate measures---for a critical study and comparison of some of these measures, see Ref.~\cite{c1}.

On a related note, since correlations and how they are distributed between a system and its embedding environment play a crucial role in the Markovianity property of the resultant system dynamics, it seems natural that tracking correlations may provide a hint for (at least partially) identifying Markovianity property of the dynamics. Along this line, here we employ a recently introduced quantum correlation measure---``discord" \cite{Zurek}---and its associated properties to propose a (non-)Markovianity measure. We illustrate the main idea through two examples.

\section{Quantum discord and Markovianity}

Quantum discord (QD) is a measure for quantumness of correlations in a bipartite state $\varrho_{SE}$ \cite{Zurek,Modi:PRL,Vedral:review}, and is defined as
\begin{eqnarray}
\mathcal{D}_S[\varrho_{SE}]=\mathcal{S}[\varrho_S]+\min_{\{\Pi_j^S\}}\mathcal{S}\left[E|\{\Pi_j^{S}\}\right] -\mathcal{S}[\varrho_{SE}],
\end{eqnarray}
where $\mathcal{S}$ is the von Neumann entropy, and $\mathcal{S}\bigl[E|\{\Pi_j^{S}\}\bigr]=\sum_j p_j \mathcal{S}\bigl[\varrho_{E|\Pi_j^S}\bigr]$ is the average entropy of the state of $E$ provided that some (rank-$1$) projective measurements $\{\Pi_j^{S}\}$ have been performed on $S$ (whence ``$S$-discord" $\mathcal{D}_{S}$) and have given the result $j$ with probability $p_j$.

\begin{figure}[tp]
\includegraphics[width=8.6cm, height=4.7cm]{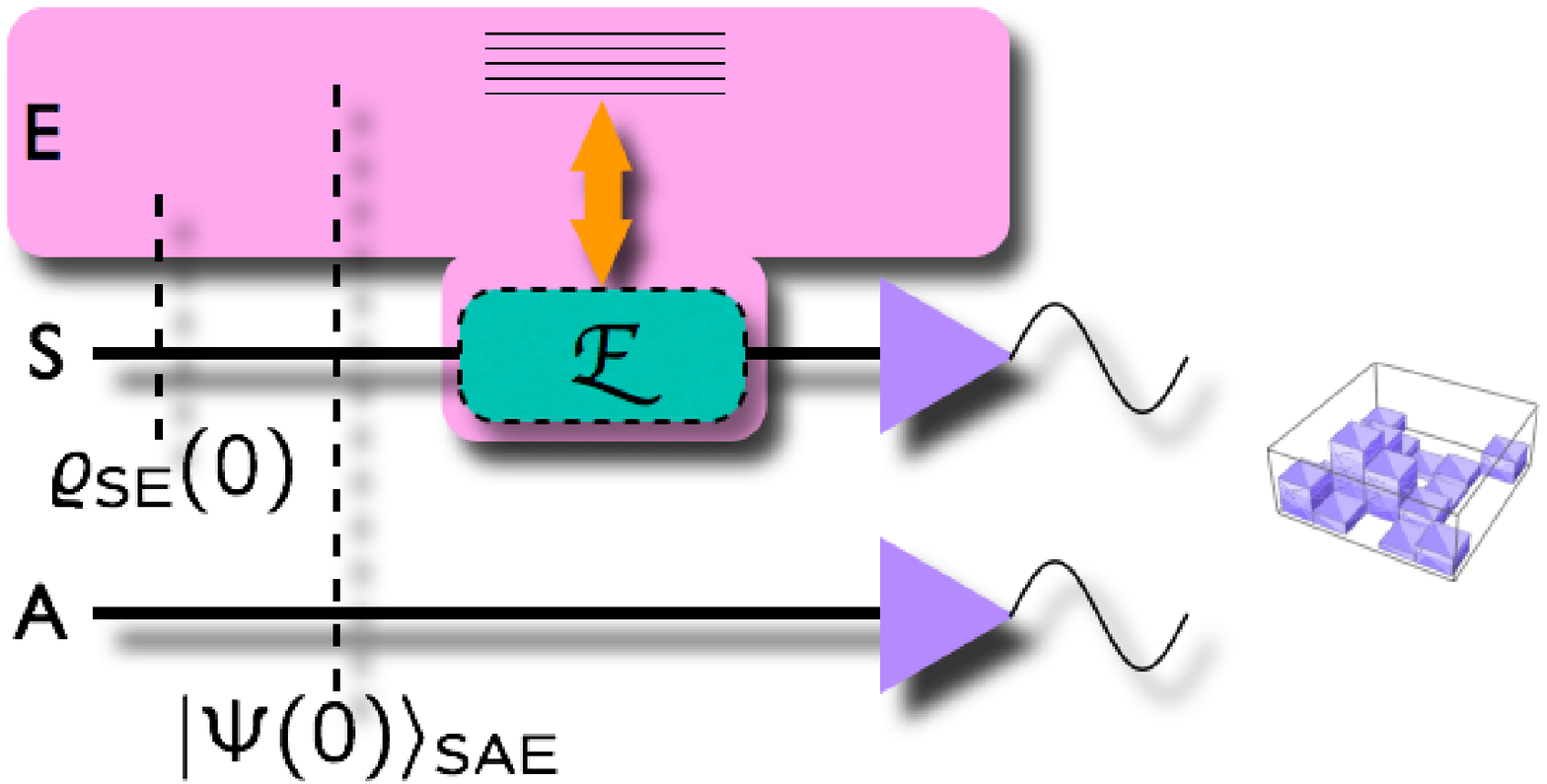}
\caption{(Color online) The initial state of the system ($S$) and environment ($E$) is a generally mixed state with vanishing discord. The state of the ancilla ($A$) is chosen such that it purifies $\varrho_{SE}(0)$; i.e, $\varrho_{SEA}(0)=|\psi(0)\rangle_{SEA}\langle \psi(0)|$. The state of the combined system ($SEA$) hence will remain pure in time. After state tomography to determine $\varrho_{\text{SA}}(t)$, we are able to conclude whether the map $\mathcal{E}$ (acting only on the system) is Markovian or non-Markovian.}
\label{discord}
\end{figure}
QD has recently attracted much attention due to its importance in distinguishing classical and quantum correlations \cite{Vedral:review,Acin,Oh, recent}, and in particular in showing how a one-qubit quantum computer may work \cite{Datta}. One can also find operational interpretations for QD, e.g., by exploiting the concept of quantum state merging \cite{QD-operational}. Vanishing and time-invariant discord in quantum states have been investigated in, respectively, Refs.~\cite{Vedral-0discord} and Ref.~\cite{Manis}.    

Of particular importance to our goal in this paper, we remind that in Ref.~\cite{Shabani}, it has been shown that, except for a ``zero-measure" set of quantum states, vanishing quantum $S$-discord of the initial state of the system and the environment is a \textit{necessary and sufficient} condition for the dynamics of the system to be CP.

Given a system-environment interaction $H_{SE}(t)$, whereby the evolution $U(t_1,t_2)=\mathrm{Texp}\bigl[-i \int_{t_1}^{t_2}H_{SE}(t)dt\bigr]$, one can define as below a \textit{CP map} $\widehat{\mathcal{E}}(t+\delta,t)$ on $\varrho_S(t)$ if(f) $\varrho_{SE}(t)$ has vanishing $S$-discord (up to the limitations stated in Ref.~\cite{Shabani}),
\begin{eqnarray}
\varrho_{S}(t+\delta)&=&\mathrm{Tr}_{E}\bigl[U_{SE}(t+\delta,t)\varrho_{SE}(t) U^{\dag}_{SE}(t+\delta,t)\bigr]\nonumber\\
&\equiv& \widehat{\mathcal{E}}(t+\delta,t)\varrho_S(t),
\end{eqnarray}
for a given $t,\delta\geq0$. One can constitute a concatenation of these maps as $\widehat{\Pi}\equiv \widehat{\mathcal{E}}(T,T-\delta) \ldots\widehat{\mathcal{E}}(t+\delta,t)\ldots \widehat{\mathcal{E}}(\delta,0)$, which by construction is an ``\textit{infinitesimal divisible}" CP map. Such maps have been proven to be always represented by a Markovian dynamics \cite{Cirac:Dividing quantum channels}. It is evident that $\widehat{\Pi}=\widehat{\mathcal{E}}(T,0)$. Thus in summary, we have the following result:

\textbf{Proposition 1:} The dynamics of a system $S$ interacting with an environment $E$ is Markovian if(f) the instantaneous combined state of the system and the environment has vanishing QD, i.e., $\mathcal{D}_S\left[\varrho_{SE}(t)\right]=0~\forall t\geq0$.

This result, as is, provides a method which is hardly experimentally measurable because it requires the knowledge of $\varrho_{SE}(t)$ at all times, which is too demanding. To overcome this problem, we relax its strict condition and just focus on the sufficiency part. Specifically, we use a lower bound for QD which is independent of the information of the environment. To this end, we use a monogamy relation for QD in a \textit{pure} tripartite state $\varrho_{SEA}$ as
\begin{eqnarray}
E_f\bigl[\varrho_{EA}\bigr]=\mathcal{D}_{S}\bigl[\varrho_{ES}\bigr]+\mathcal{S}_{E|S},
\label{discord-conservation}
\end{eqnarray}
where $E_f\bigl[\varrho_{EA}\bigr]$ is the entanglement of formation of the state of $EA$ \cite{EoF}, and $\mathcal{S}_{E|S}=\mathcal{S}\bigl[\varrho_{ES}\bigr]-\mathcal{S}\bigl[\varrho_{S}\bigr]$ is the conditional entropy \cite{de Oliveira} . Since the total state $\varrho_{SEA}$ is assumed to be pure, from the Schmidt decomposition we have $\mathcal{S}\bigl[\varrho_{ES}\bigr]=\mathcal{S}\bigl[\varrho_{A}\bigr]$, whence
\begin{eqnarray}\label{QD-monogamy}
E_f\bigl[\varrho_{EA}\bigr]=\mathcal{D}_{S}\bigl[\varrho_{ES}\bigr] + \mathcal{S}\bigl[\varrho_{A}\bigr]-\mathcal{S}\bigl[\varrho_{S}\bigr].
\label{discord-conservation-1}
\end{eqnarray}

Now let us purify the system-environment initial state $\varrho_{SE}(0)$ by attaching an ancilla $A$ which is not interacting with $S$ and $E$. The total state $\varrho_{SEA}(t)$ will remain pure in time because $SE$ evolves unitarily in time, while $A$ does not evolve. In this case, noting that entanglement of formation is always nonnegative, Eq.~(\ref{discord-conservation-1}) yields
\begin{eqnarray}
\label{discord-lowerbound}
\mathcal{D}_{S}\bigl[\varrho_{ES}(t)\bigr]\geq \mathcal{S}\bigl[\varrho_{S}(t)\bigr] - \mathcal{S}\bigl[\varrho_{A}(t)\bigr]=:\Delta_{SA}(t).
\end{eqnarray}
This relation implies that to recognize non-Markovianity of a quantum dynamics, it suffices to measure $\Delta_{SA}(t)$; an strictly positive value for $\Delta_{SA}$ at some time $t_*$ is a signature of non-Markovianity of the dynamics at any later time.

Proposition 1 can also be used to define a measure for non-Markovianity power of a quantum dynamics in an interval $(0,\tau)$ as follows:
\begin{eqnarray}
P_{\text{NM}}(\tau)=\frac{1}{\tau}\int_0^{\tau} \mathcal{D}_{S}\bigl[\varrho_{SE}(t)\bigr] dt.
\end{eqnarray}
A nonzero value for $P_{\text{NM}}$ is a necessary and sufficient condition for non-Markovianity of the associated quantum dynamics. Since again computing $\mathcal{D}_{S}\bigl[\varrho_{SE}(t)\bigr]$ is difficult, we propose
\begin{eqnarray}
\widetilde{P}_{\text{NM}}(\tau)=\frac{1}{2\tau}\int_0^{\tau} \big(|\Delta_{SA}(t)|+\Delta_{SA}(t)\big) dt,
\label{non-Markovian-power-lowerbound}
\end{eqnarray}
as a lower bound for non-Markovianity power.

\textit{Remark 1.} It should be noted that our QD measure is not a trivial extension of the entanglement measure for non-Markovianity power of a quantum dynamics. One of the characteristics of QD is that it can increase under local operation and classical communication (LOCC) while entanglement cannot. Separable states $\varrho_{SE}=\sum_i p_i \varrho_{S}^{(i)}\otimes \varrho_{E}^{(i)}$ \big($0\leq p_i \leq 1$, $\sum_{i}p_i=1$\big) can be generated by LOCC from some initial product state, say, $\varrho_{SE}(0)=\ketbra{0}{0}\otimes\ketbra{1}{1}$. Since there exist separable states with nonvanishing QD \cite{Zurek}, LOCC may increase QD. As an explicit example, consider the preparation of the system and ancilla in the vanishing QD state $\varrho_{SA}(0)=\frac{1}{2}(\ketbra{0}{0}\otimes w_1+\ketbra{1}{1}\otimes w_2)$, where $w_1$ and $w_2$ are two orthogonal density matrices. A Hadamard channel
\begin{eqnarray}
\widehat{\mathcal{E}}(\tau)\varrho=[1-p(\tau)]\varrho+p(\tau)H \varrho H,
\end{eqnarray}
with $H=(\sigma^{(1)} + \sigma^{(3)})/\sqrt{2}$, acting on the system yields
\begin{eqnarray}
\varrho_{SA}(\tau)&=&\bigl(\widehat{\mathcal{E}}(\tau)\otimes\openone\bigr)\varrho_{SA}(0)\nonumber\\
&=&\frac{1}{2}\{[1-p(\tau)]\ketbra{0}{0}+p\ketbra{+}{+}\}\otimes w_1+\nonumber\\&&~\frac{1}2 \{[1-p(\tau)]\ketbra{1}{1}+p(\tau)\ketbra{-}{-}\}\otimes w_2,
\end{eqnarray}
in which $\ket{\pm}$ is the eigenvector of $\sigma^{(1)}$. Since $[\varrho_{SA}(\tau),\varrho_{S}(\tau)\otimes \openone]\neq 0$, $\varrho_{SA}(\tau)$ is a state with nonvanishing QD \cite{Acin}, while $\mathcal{D}_S[\varrho_{SA}(0)]=0$.

\textit{Remark 2.} The positivity of the associated Choi-Jamilkowski state of a quantum process has been used as a measure for its Markovianity \cite{Rivas}. To use this criterion, one needs to find $\widehat{\mathcal{E}}(t+\epsilon,t)$. But the problem is that in general $\widehat{\mathcal{E}}(t+\epsilon,t)$ is not a well-defined quantum process/map. When the evolution is non-Markovian, $\widehat{\mathcal{E}}(t+\epsilon,t)$ is not necessarily a map or identifiable by a physically conceivable tomography scheme. In fact, in the middle of the process at some time $t_*$, it may happen that $\varrho_{SE}(t_*)$ has nonzero quantum correlation. Hence, the evolution of the system cannot be interpreted as the application of a quantum \textit{map} on $\varrho_{S}$. Note also that the number of independent parameters to be evaluated to identify the process between a $t$ and $t+\epsilon$ thoroughly is $O(d^4)$, where $d$ is the dimension of the Hilbert space of the system; while a complete set of quantum states has only $O(d^2)$ independent members. This implies that process tomography for identification of $\widehat{\mathcal{E}}(t+\epsilon,t)$ is in general infeasible. In our approach, however, all one needs is just state tomography of $\varrho_{SE}(t)$ in every point of time, which is well-defined (although of course difficult). Besides, if one use the lower bound proposed, our measure seems experimentally feasible.

\textit{Remark 3.} It should be clear that, although we use the result of Ref.~\cite{Shabani}, here we have set different goals and strategy. We deal with the problem of non-Markovianity and provide a measure thereof by using the concept of QD, while Ref.~\cite{Shabani} deals with a condition for CPT-ness of quantum maps.

\begin{figure}[tp]
\includegraphics[width=9cm, height=6.2cm]{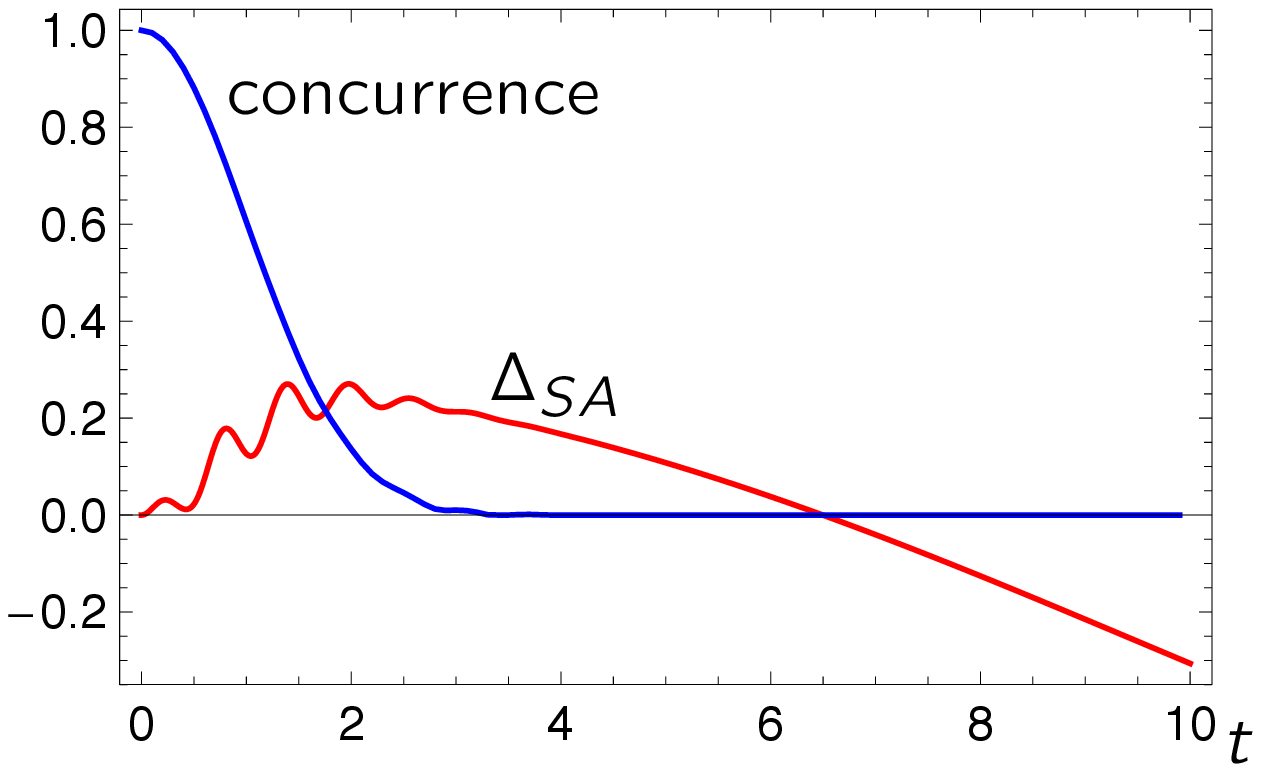}
\vskip-9mm
\caption{(Color online) Concurrence \cite{EoF} (blue line) of $\varrho_{SA}$ is decreasing in every arbitrary time interval [$t$ is in $\hbar\equiv1$ units]. Hence, deciding about Markovianity of the dynamics is not possible through this entanglement measure. On the other hand, $\Delta_{SA}$ [Eq.~(\ref{discord-lowerbound})] (the red curve), is positive in $t\in[0,6]$, which shows that $\mathcal{D}_{S}\bigl[\varrho_{SE}(t)\bigr]$ is nonzero in this interval; hence, the process is non-Markovian.}
\label{entropyvsnegativity}
\end{figure}

\section{Examples}

In the following, we illustrate the main idea of this paper through two examples.

\textbf{Example I.} Consider the Jaynes-Cummings model, in which the system of interest is a two-level atom interacting with a cavity mode through the Hamiltonian
\begin{eqnarray}
H_{SE}= \lambda (\sigma^{-} \otimes a^\dag +\sigma^{+}\otimes a),
\end{eqnarray}
where $\sigma^+=\ketbra{1}{0}$ and $a^\dag$ are, respectively, the raising operators of the system and environment. We assume the initial state of the atom ($S$) and cavity ($E$) to be a separable state $\varrho_{SE}(0)=\varrho_{S}(0)\otimes |\alpha\rangle_E\langle \alpha|$, in which $\ket{\alpha}$ is a coherent state, and $\lambda=1$. For the case of $\varrho_{S}(0)=\epsilon \ketbra{0}{0}+(1-\epsilon) \ketbra{1}{1}$,  the purification with a qubit ancilla yields $|\psi(0)\rangle_{SAE}=(\sqrt{\epsilon} \ket{00}+\sqrt{1-\epsilon}\ket{11})\otimes \ket{\alpha}$. When $\epsilon=0.2$ and $\alpha=5$, the concurrence of the system and ancilla decreases in time. Hence, the entanglement measure of non-Markovianity does not imply a reliable clue about Markovianity/non-Markovianity of the dynamics. But since $\Delta_{SA}(t)>0$, we are sure that the evolution is non-Markovian (see Fig.~\ref{entropyvsnegativity}). Calculation of the lower bound on non-Markovianity power of the dynamics gives $\widetilde{P}_{\text{NM}}(10)= 0.094$.

\begin{figure}[tp]
\vskip-4mm
\includegraphics[width=11cm, height=7.3cm]{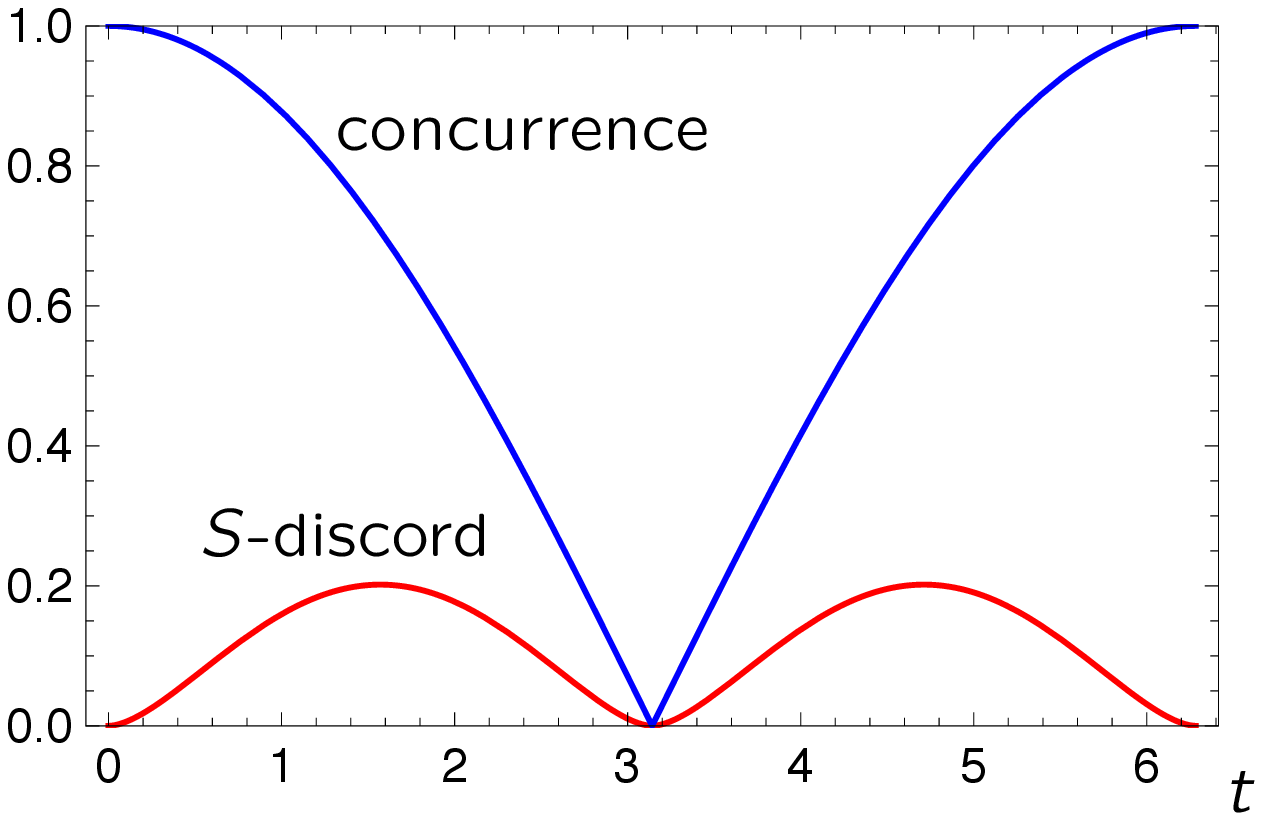}
\vskip-16mm
\caption{(Color online) $S$-discord (red) of example II vs. time [in $\hbar\equiv1$ units] for $\varrho_{SE}(0)=\ketbra{\phi}{\phi}\otimes (p\ketbra{0}{0}+(1-p)\ketbra{1}{1})$, where $\ket{\phi}=(\ket{0}+\ket{1})/\sqrt{2}$ and $p=0.5$. $S$-discord has been obtained from Eq.~(\ref{discord-conservation-1}) in which $\ket{\psi(t)}_{SEA}=(U_{SE}(t)\otimes \openone)[\ket{\phi}\otimes(\sqrt{p}\ket{00}_{EA}+\sqrt{1-p}\ket{11}_{EA})]$. The blue curve shows concurrence between the system and ancilla vs. time. Since concurrence increases only after $t_*=3$, the non-Markovianity property of the dynamics for $t<t_*$ is not captured by using the entanglement measure of non-Markovianity.}
\label{discord-phase-flip-channel}
\end{figure}
\textbf{Example II.} Consider two qubits (system-environment) interacting through the Hamiltonian
\begin{eqnarray}
H=\frac{1}{4}(\openone+\sigma_{1}^{(3)}+\sigma_{2}^{(3)}-\sigma_{1}^{(3)}\sigma_{2}^{(3)}),
\end{eqnarray}
leading to the evolution operator
\begin{eqnarray}
U_{SE}(t)=\sum_{j\in\{0,3\}} e^{-i t\sigma^{(j)}/2}\otimes \ket{j}\bra{j},
\end{eqnarray}
 in which $(\sigma^{(0)},\sigma^{(1)},\sigma^{(2)},\sigma^{(3)})=(\openone,X,Y,Z)$ are the identity and the Pauli matrices.

We show that if the initial state of the environment is diagonal in the computational basis, the state of the environment does not change at any time, $\varrho_{E}(t)=\varrho_{E}(0)$, and the system and the environment remain unentangled in time---see, e.g., Ref. \cite{LoF} for a note on lack of back-action in revival of quantum correlations. One may infer that the evolution of the system should then be described by a quantum dynamical semigroup. However, we show that it is a non-Markovian transformation because a nonzero QD is produced during the evolution due to the interaction of the system and the environment. Assuming $\varrho_{SE}(0)=\varrho_{S}(0)\otimes\varrho_{E}(0)$, the Kraus operators of the dynamics of the system are $A_0=\sqrt{p_0}~\openone$ and $A_1= \sqrt{p_1}~e^{-it \sigma^{(3)}/2}$, where $p_j=\bra{j}\varrho_{E}\ket{j}$. To find whether the obtained time-dependent map $\widehat{\mathcal{E}}(\tau)~\cdot~=\sum_j A_j(\tau)\cdot A_j^{\dag}(\tau)$ is Markovian or not, one needs to calculate the $S$-discord of $\varrho_{SE}(t)$. It is straightforward to show that
\begin{eqnarray}
\hskip-2mm\varrho_{SE}(t)= \sum_{j\in\{0,3\}} p_j~e^{-it \sigma^{(j)}/2}\varrho_S(0)~e^{it \sigma^{(j)}/2} \otimes \ket{j}\bra{j},
\label{ex1}
\end{eqnarray}
which is a separable state. Given the environment initially in the state $\varrho_{E}(0)=p_0 \ketbra{0}{0}+p_1 \ketbra{1}{1}$, it is evident that $\varrho_{E}(t)=\varrho_{E}(0)$.

Now we use a result of Ref.~\cite{Acin} that if a bipartite quantum state has zero $S$-discord, then $\bigl[\varrho_{SE},\varrho_{S}\otimes \openone\bigr]= 0$. Since in this example $[\varrho_{SE}(t),\varrho_{S}(t)\otimes \openone]\neq 0$ unless $[\varrho_{S}(0),\sigma^{(3)}]=0$, thus except for these special states, $S$-discord is nonzero, and hence the dynamics of the system is non-Markovian (see Fig.~\ref{discord-phase-flip-channel}).

\section{Summary and outlook}

An interesting question in open quantum system theory is that given the interaction Hamiltonian of a system and its environment whether the induced dynamics on the system is Markovian or not. Several methods have been suggested to assess non-Markovianity of a dynamics. Here, based on the idea that underlying correlations play key role in a reduced dynamics, we employ quantum discord (QD) to define a measure for non-Markovianity. To alleviate the need for the state of the environment in calculating QD in our scheme, we have considered an experimentally measurable lower bound for QD by using a monogamy property for tripartite quantum systems. This bound being nonvanishing provides a sufficient condition for non-Markovianity of a quantum dynamics.

We also have discussed two examples to illustrate the result of this study. In the first example, it has been shown that QD can signal non-Markovianity in a process which eludes the entanglement measure of non-Markovianity. The second example has featured an explicit quantum dynamics whose non-Markovianity is strictly due to the production of quantum correlation not in the form of entanglement. Here, the initial state of the system and environment was a product state and remained separable in time, while the state of the environment did not vary in time either.

Having powerful non-Markovianity measures at hand, one can think of various applications in quantum information theory. For example, an intriguing question is that how one should engineer an environment so that the generated sub-dynamics of the attached system becomes Markovian. This is important because, for example, it has been argued \cite{Verstraete-engineering} that one can perform adiabatic quantum computation in an indirect fashion in an open system, provided that the effective system dynamics is a Markovian evolution with fixed points encoding ground states of some given Hamiltonian. Such scenarios require ability to decide whether an environment induces Markovian sub-dynamics. Our method provides a way to go to answer this question. Given an interaction Hamiltonian with unknown coupling constants, $\lambda$, and a preparation of the system and environment (assuming being feasible), $\varrho_{SE}(0)$, we can find the state of the system and the environment as a function of $\lambda$, $\varrho_{SE}(\lambda, t)$, and check for which values of  $\lambda$, $S$-discord remains zero in time. As a result, we are in principle able to manipulate coupling constants such that the desired map becomes Markovian or non-Markovian on demand.

\textit{Acknowledgements.---} The authors acknowledge V. Karimipour, D. A. Lidar, and A. Shabani for useful discussions. An earlier version of this work was presented in the School and Workshop of New Trends in Quantum Dynamics and Entanglement (the Abdus Salam International Center for Theoretical Physics, Trieste, Italy, 2011); we appreciate all comments by the participants of this school/workshop. This research is partially supported by Sharif University of Technology's office of vice-chair for research.



\begin{thebibliography}{99}

\bibitem{Breuerbook} H.-P. Breuer and F. Petruccione, \textit{The Theory of Open Quantum Systems} (Oxford University Press, New York, 2002); H. Carmichael, \textit{An Open System Approach to Quantum Optics} (Springer, Berlin, 1994); A. Rivas and S. F. Huelga, arXiv:1104.5242.

\bibitem{Cubitt-2} T. S. Cubitt, J. Eisert, and M. M. Wolf, arXiv:1005.0005.

\bibitem{Verstraete-engineering} F. Verstraete, M. M. Wolf, and J. I. Cirac, Nature Phys. \textbf{5}, 633 (2009).

\bibitem{Lidar-chem} D. A. Lidar, Z. Bihary, and K. B. Whaley, Chem. Phys. \textbf{268}, 35 (2001).

\bibitem{Nielsen:book} M. A. Nielsen and I. L. Chuang, \textit{Quantum Computation and Quantum Information} (Cambridge University Press, Cambridge, England, 2000).

\bibitem{cirac:Assessing NM} M. M. Wolf, J. Eisert, T. S. Cubitt, and J. I. Cirac, Phys. Rev. Lett. \textbf{101}, 150402 (2008).

\bibitem{Andersson} E. Andersson, J. D. Cresser, and M. J. W. Hall, arXiv:1009.0845.

\bibitem{Cirac:Dividing quantum channels} M. M. Wolf and J. I. Cirac, Commun. Math. Phys. \textbf{279}, 147 (2008).

\bibitem{Lindblad}
V. Gorini, A. Kossakowski, and E. C. G. Sudarshan, J. Math. Phys. \textbf{17}, 821 (1976); G. Lindblad, Commun. Math. Phys. \textbf{48}, 119 (1976).

\bibitem{c2} A. Rivas, A. D. K. Plato, S. F. Huelga, and M. B. Plenio, New J. Phys. \textbf{12}, 113032 (2010).

\bibitem{Ishizaki-Fleming:09} A. Ishizaki and G. R. Fleming, J. Chem. Phys. \textbf{130}, 234110 (2009);
\textit{ibid.} \textbf{130}, 234111 (2009).

\bibitem{Cheng-Fleming:09} Y.-C. Cheng and G. R. Fleming, Annu. Rev. Phys. Chem. \textbf{60}, 241 (2009).

\bibitem{SL-pM} A. Shabani and D. A. Lidar, Phys. Rev. A \textbf{71}, 020101(R) (2005);
H.-P. Breuer and B. Vacchini, Phys. Rev. Lett. \textbf{101}, 140402 (2008).

\bibitem{NM-papers} H.-P. Breuer, J. Gemmer, and M. Michel, Phys. Rev. E \textbf{73}, 016139 (2006);
H.-P. Breuer, Phys. Rev. A \textbf{75}, 022103 (2007).

\bibitem{Cubitt} T. S. Cubitt, J. Eisert, and M. M. Wolf, Commun. Math. Phys. \textbf{310}, 383 (2012).

\bibitem{Breuer} H.-P. Breuer, E.-M. Laine, and J. Piilo, Phys. Rev. Lett. \textbf{103}, 210401 (2009).

\bibitem{indian} A. K. Rajagopal, A. R. Usha Devi, and R. W. Rendell, Phys. Rev. A \textbf{82}, 042107 (2010).

\bibitem{Rivas} A. Rivas, S. F. Huelga, and M. B. Plenio, Phys. Rev. Lett. \textbf{105}, 050403 (2010).

\bibitem{c1} 
B. Vacchini, A. Smirne, E.-M. Laine, J. Piilo, and H.-P. Breuer,  New J. Phys. \textbf{13}, 093004 (2011)

\bibitem{Zurek} H. Ollivier and W. H. Zurek, Phys. Rev. Lett. \textbf{88}, 017901 (2002);
L. Henderson and V. Vedral, J. Phys. A: Math. Gen. \textbf{34}, 6899 (2001).

\bibitem{Modi:PRL} K. Modi, T. Paterek, W. Son, V. Vedral, and M. Williamson, Phys. Rev. Lett. \textbf{104}, 080501 (2010).

\bibitem{Vedral:review} K. Modi, A. Brodutch, H. Cable, T. Paterek, and V. Vedral, arXiv:1112.6238.

\bibitem{Acin} A. Ferraro, L. Aolita, D. Cavalcanti, F. M. Cucchietti, and A. Ac\'{\i}n, Phys. Rev. A \textbf{81}, 052318 (2010).

\bibitem{Oh}  C. Zhang, S. Yu, Q. Chen, and C. H. Oh, Phys. Rev. A \textbf{84}, 032122 (2011);
\textit{ibid.} 052112 (2011);
Q. Chen, C. Zhang, S. Yu, X. X. Yi, and C. H. Oh, \textit{ibid.} 042313 (2011).

\bibitem{recent}M. Arsenijevi\'{c}, J. Jekni\'{c}-Dugi\'{c}, and M. Dugi\'{c}, arXiv:1203.4612.
%

\bibitem{Datta} A. Datta, A. Shaji, and C. M. Caves, Phys. Rev. Lett. \textbf{100}, 050502 (2008).

\bibitem{QD-operational} V. Madhok and A. Datta, Phys. Rev. A \textbf{83}, 032323 (2011);
D. Cavalcanti, L. Aolita, S. Boixo, K. Modi, M. Piani, and A. Winter, \textit{ibid.}, 032324 (2011).

\bibitem{Vedral-0discord} B. Daki\'{c}, V. Vedral, and \v{C}. Brukner, Phys. Rev. Lett. \textbf{105}, 190502 (2010);
B. Bylicka and D. Chru\'{s}ci\'{n}ski, Phys. Rev. A \textbf{81}, 062102 (2010); 
B. Bylicka and D. Chru\'{s}ci\'{n}ski, arXiv:1104.1804. 

\bibitem{Manis} L. Mazzola, J. Piilo, and S. Maniscalco, Intl. J. Quantum Inf. \textbf{9}, 981 (2011). 

\bibitem{Shabani} A. Shabani and D. A. Lidar, Phys. Rev. Lett. \textbf{102}, 100402 (2009).

\bibitem{EoF} R. Horodecki, P. Horodecki, M. Horodecki, and K. Horodecki, Rev. Mod. Phys. \textbf{81}, 865 (2009).

\bibitem{de Oliveira} F. F. Fanchini, M. F. Cornelio, M. C. de Oliveira, and A. O. Caldeira, Phys. Rev. A \textbf{84}, 012313 (2011).

\bibitem{LoF} R. Lo Franco, B. Bellomo, E. Andersson, and G. Compagno, Phys. Rev. A \textbf{85}, 032318 (2012).

\end{thebibliography}
\end{document}